# Forced Oscillation Source Location via Multivariate Time Series Classification

Yao Meng[1,2], Zhe Yu[1], Di Shi[1], Desong Bian[1], Zhiwei Wang[1]

[1]GEIRI North America, San Jose, CA 95134
[2]North Carolina State University, Raleigh, NC
Emails: ymeng3@ncsu.edu; {zhe.yu, di.shi, desong.bian, zhiwei.wang}@geirina.net

*Abstract*—Precisely locating low-frequency oscillation sources is the prerequisite of suppressing sustained oscillation, which is an essential guarantee for the secure and stable operation of power grids. Using synchrophasor measurements, a machine learning method is proposed to locate the source of forced oscillation in power systems. Rotor angle and active power of each power plant are utilized to construct multivariate time series (MTS). Applying Mahalanobis distance metric and dynamic time warping, the distance between MTS with different phases or lengths can be appropriately measured. The obtained distance metric, representing characteristics during the transient phase of forced oscillation under different disturbance sources, is used for offline classifier training and online matching to locate the disturbance source. Simulation results using the four-machine two-area system and IEEE 39-bus system indicate that the proposed location method can identify the power system forced oscillation source online with high accuracy.

## I. Introduction

Low-frequency oscillation reduces power transmission limit and may lead to damage of system equipment. It severely threatens the security and stability of large-scale interconnected power systems in real-time operations. Insufficient damping of a system results in a majority of low-frequency oscillation, which can be suppressed by tuning parameters of power system stabilizer or intertie line controls. However, forced oscillations caused by resonance have been discovered recently [1] in power grids. The posterior analysis shows the system was well damped when the oscillation occurred. Moreover, disturbance with a frequency approximating the intrinsic system frequency was injected into the power system somewhere. The resonance excited and even the small disturbance could amplify and spread rapidly in the whole power system. The traditional remediation actions, for example putting power system stabilizers into operation, etc., are not applicable for suppressing such oscillations. The most effective way to quench forced oscillations is to remove the disturbance rapidly and accurately.

Locating the oscillation source is the prerequisite to eliminating the disturbance. With massive Phasor Measurement Units (PMUs) installed, monitoring dynamic behaviors of the power system has become possible. Taking advantage of PMU measurements, the research community has proposed several forced oscillation source location methods. Integrating PMU measurements at different places and the wave speed map, a traveling wave based method is proposed in [2]. Based on the transient energy function, the energy flow direction can be calculated to locate the forced oscillation source, which has good location performance [3]-[5]. Authors of [6] locate the oscillation source by estimating the mode shape, which represents the relative magnitude and phasing of the oscillation throughout the system. Using tie-line power signal during oscillations, the real-time approximate entropy value in a continuous time interval, which is corresponding to the location of disturbance in power systems, can be calculated [7]. In [8], a machine learning approach is proposed. Measurement signals during forced oscillations are mapped to a multi-dimensional CELL, and decision tree is utilized to identify the characteristic parameters of the CELL corresponding to different oscillation sources.

The discoveries and investigations of these methods provide a better understanding of applying PMU measurements to locate the disturbance source. Nonetheless, some approaches suffer from a lack of theoretical support. Moreover, most literature assumes the low-frequency oscillation can be immediately detected after it occurs and utilizes the information at the beginning phase of oscillations, which are not practical in general.

A machine learning based method to locate the forced oscillation sources is proposed in this paper. Rotor angle and active power of each generation measured by PMUs are utilized to construct multivariate time series (MTS). Mahalanobis distance and dynamic time warping (DTW) are applied to represent the distance between out-of-sync MTS. With the obtained distance metric, offline training and online matching can be conducted. In this approach, MTS with different length can be appropriately compared which relaxes the assumption of accurate detection of the beginning of oscillations. Case studies on the four-machine two-area system and IEEE 39-bus system indicate the effectiveness of the proposed approach. In addition, the effect of time delay in oscillation detection on locating accuracy is discussed.

This work is funded by SGCC Science and Technology Program under contract No. 5455HJ160007.

## II. MULTIVARIATE TIME SERIES CLASSIFICATION

A machine learning approach that utilizes PMU measurements of generator rotor angle and active power in power grids to locate the source of forced oscillation geographically down to the substation level is proposed. It is assumed that at least one PMU measurement is available for each power plant, and the model of the power system is also available.

Fig.1 illustrates the proposed approach procedure. When forced oscillation occurs, the response of generators tends to be distinct for different oscillation source locations. Based on this intuition, dynamics of generators under various scenarios of forced oscillation source locations are generated through time domain simulation. Simulated PMU measurements of rotor angles and active power of generators are utilized to construct signature multivariate time series. Mahalanobis distance is trained offline through metric learning. When a forced oscillation is detected, the same coefficients are measured and compared with the trained classifier to determine the source location of the oscillation. Dynamic time warping is applied to handle the out-of-sync between testing data sets and training data sets due to the oscillation detection delay.

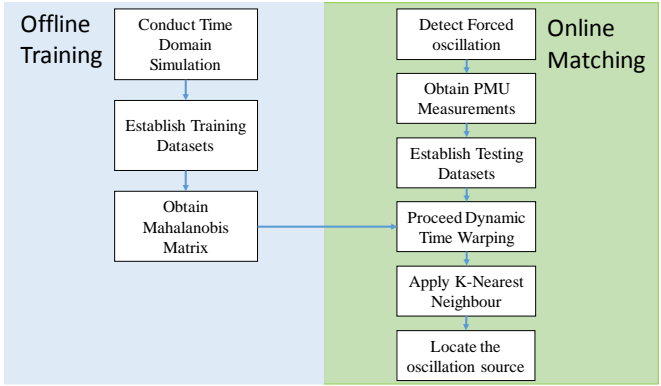

Fig.1 Flowchart of forced oscillation source identification based on multivariate time series classification.

### A. Mahalanobis Distance

Mahalanobis Distance is a standard measure of the distance, which is characterized by a symmetry Positive Semi-Definite (PSD) matrix $M$. For two vectors $x$ and $y$, where $x, y \in R^d$, the squared Mahalanobis distance between $x$ and $y$ is defined as follows:

$$d_M(x, y) = (x - y)^T M (x - y) \quad (1)$$

If $M=I$, Mahalanobis distance degenerates to the standard Euclidean distance.

Since Mahalanobis distance considers correlation among different variables, an accurate relationship between variables and labels of MTS can be established. It has two essential functions. The first one is to remove correlations among different variates and to map the original space into a new coordinate system. The second one is to assign weights to new variates [9]. With these two functions, Mahalanobis distance can measure the distance between vectors effectively.

Next, the Mahalanobis metric will be extended to measure the distance between two multivariate time series. Given two MTS $X$ and $Y$,

$$X = \begin{bmatrix} x_1(1) & x_2(1) & \cdots & x_p(1) \\ x_1(2) & x_2(2) & \cdots & x_p(2) \\ \vdots & \vdots & \ddots & \vdots \\ x_1(h) & x_2(h) & \cdots & x_p(h) \end{bmatrix} \quad (2)$$

and

$$Y = \begin{bmatrix} y_1(1) & y_2(1) & \cdots & y_p(1) \\ y_1(2) & y_2(2) & \cdots & y_p(2) \\ \vdots & \vdots & \ddots & \vdots \\ y_1(h) & y_2(h) & \cdots & y_p(h) \end{bmatrix} \quad (3)$$

where $p$ is the number of features, $h$ is the number of sampling points. The local distance measure is expressed as

$$d_M(X^i, Y^j) = (X^i - Y^j) M (X^i - Y^j)^T \quad (4)$$

where $X^i$ represents the $i^{th}$ row in $X$, and $Y^j$ represents the $j^{th}$ row in $Y$. Then the distance between multivariate time series $X$ and $Y$ is defined as

$$D_M(X, Y) = \sum_{j=1}^{h} d_M(X^j, Y^j) \quad (5)$$

### B. Metric Learning

The Mahalanobis metric represents the relevance of two time series. The goal of metric learning is utilizing the labeled training data to find an appropriate $M$ such that the Mahalanobis distance emphasizes the relevant features while decreases the effect of irrelevant features [10].

For any given triplet label {$X, Y, Z$}, where $X$ and $Y$ are in the same class while $Z$ is in a different one, the Mahalanobis distance between instance $X$ and instance $Y$ should be smaller than the distance between $X$ and $Z$.

The metric learning problem can be formulated as an optimization problem as follows:

$$\begin{aligned} & \min_{M} 1 \\ & s.t.\ D_M(X, Y) - D_M(X, Z) < -\rho,\ \forall \{X, Y, Z\} \\ & M \in \text{PSD} \end{aligned} \quad (6)$$

Here, $\rho>0$ denotes the desired margin. The objective of (6) is to find a PSD matrix to satisfy all triplet constraints {$X, Y, Z$}. The number of triplet label [11] constraints is the cubic of the number of the training samples. In [12], the authors have proposed a strategy to choose the most useful triplets for training.

To solve the optimization proposed in (6), an iterative process is proposed, which can be written as:

**Algorithm 1 Train Mahalanobis Matrix through Metric Learning**
1:  Initialize Mahalanobis matrix $M_0$

2:     For *i=1:MAX₁*
3:         For *j=1:MAX₂*
4:             Choose the most useful triplets constraint {$X_j$, $Y_j$, $Z_j$}
5:             If constraint (6) is violated, calculate loss function $l(M_j)$
6:             Update Mahalanobis matrix $M_j$
7:         End for
8:         Calculate total loss function $L_i=\sum_j l(M_j)$
9:         Check whether $L_i$ is less than threshold, if yes, break
10:    End for

MAX1 and MAX2 are set to limit the number of iterations. If the triplet constraint is violated, the loss function is defined as:

$$l(M) = \rho + D_M(X_j, Y_j) - D_M(X_j, Z_j) \quad (7)$$

How to update Mahalanobis matrix is the key problem in Algorithm 1. The Mahalanobis matrix $M_j$ should be updated to reduce the value of loss function. What's more, to avoid unstable learning process, a regularization term which restricts the matrix divergence between two different iterations should be added to the metric learning objective function. So the Mahalanobis matrix updating equation can be expressed as:

$$M_{j+1} = \arg\min_{M \succ 0} div(M_j, M) + \lambda l(M) \quad (8)$$

where $\lambda$ is the regularization parameter balancing the loss function $l(M_i)$ and the regularization function $div(M_j, M_{j+1})$. The regularization function $div(M_j, M_{j+1})$ measures the matrix divergence. It can be expressed as:

$$div(M_j, M_{j+1}) = tr(M_j M_{j+1}^{-1}) - \log(\det(M_j M_{j+1}^{-1})) - n \quad (9)$$

where tr() denotes the trace of a matrix, $n$ is the dimension of $M$ [13].

We solve (8) in an iterative way. To ensure the obtained Mahalanobis matrix $M_{j+1}$ is a PSD matrix in each iteration, we require

$$\begin{cases} \lambda_j (P_j P_j^T - Q_j Q_j^T) + M_j^{-1} \geq 0 \\ \lambda_j \geq 0 \end{cases} \quad (12)$$

Several tools can solve this standard linear matrix inequalities (LMIs). If the obtained result is $\bar{\lambda}_j$, $\lambda_j \in [0, \bar{\lambda}_j]$ ensures that updated $M_{j+1}$ is a PSD matrix. Therefore, we solve LMIs first in each iteration and select $\lambda_j$ in the feasible range.

Given $\lambda_j$, equation (8) reaches its minimum when its gradient is zero. By setting the gradient of (8) to be zero, we get:

$$M_{j+1} = (M_j^{-1} + \lambda_j (P_j P_j^T - Q_j Q_j^T))^{-1} \quad (10)$$

where $P_j=X_j-Y_j$, $Q_j=X_j-Z_j$. We can solve (10) by applying Woodbury matrix identity, the iterative expression of $M_j$ is:

$$\begin{cases} \gamma_j = M_j - \lambda_j M_j P_j (I + \lambda_j P_j^T M_j P_j)^{-1} P_j^T M_j \\ M_{j+1} = \gamma_j + \lambda_j \gamma_j Q_j (I - \lambda_j Q_j^T \gamma_j Q_j)^{-1} Q_j^T \gamma_j \end{cases} \quad (11)$$

where $\gamma_j=(M_j^{-1}+\lambda_j P_j P_j^T)^{-1}$.

Through this process, Mahalanobis matrix can be updated if triplet constraint is violated. When calculated total loss function is smaller than a predefined threshold or *MAX₁* is reached, the algorithm terminates. Metric learning is the most compute-intensive part of the proposed approach, but it can be done offline, which is not a time-critical task.

### C. Dynamic Time Warping

When applying multivariate time series classification, it is unreasonable to assume the beginning and ending time points of interest can be correctly identified, especially during the later deployment. In the oscillation locating context, detection of the beginning of the forced oscillation event is not guaranteed. Thus, we need to deal with time series analysis with different phases and lengths. Dynamic Time Warping (DTW) is an algorithm which conducts nonlinear mapping of one time series to another by minimizing the distance [14]. Through calculating the optimal warp path, two time series will be extended and placed in one-to-one correspondence, which makes their similarity can be measured easily.

Given two time series, $Q(i)$, $i=1,2,...,m$ and $C(k)$, $k=1,2,...,n$, define an optimal warp path $W$ as

$$W(j) = \begin{pmatrix} w_Q(j) \\ w_C(j) \end{pmatrix}, j = 1, 2, ..., s \quad (13)$$

where $w_Q(j) \in [1, m]$ denotes the index in sequence $Q$, $w_C(j) \in [1, n]$ denotes the index in sequence $C$, and $s$ is the length of the warp path. $(w_Q(j), w_C(j))^T$ means the $w_Q(j)^{th}$ element of $Q$ and the $w_C(j)^{th}$ element of $C$ correspond to each other.

To reduce the number of paths during the search, a valid warping path should satisfy several well-known conditions. Boundary condition ensures all indices of each time series are used in the warping path. Continuity condition requires the warping path to be made of only adjacent cells. Moreover, monotonicity condition restricts the feasible warping path only increase monotonically. These three conditions can be expressed as

$$\begin{cases} W(1) = (1,1)' \\ W(s) = (m,n)' \\ D(i,k) = d(Q(i), C(k)) + \\ \quad \min(D(i-1,k-1), D(i-1,k), D(i,k-1)) \end{cases} \quad (14)$$

where $d(i,k)$ is the distance found in the current cell, $D(i,k)$ represents the minimum sub warp path distance, and the length of the warping path $s \in [\max(m, n), m+n]$. The optimal warping path can be found using dynamic programming. After all of the elements in the distance matrix $D(i,k)$ are calculated, the corresponding warping path is the optimal warping path $W$. Fig.2 illustrates an example of optimized warping path between two given time series. The shadow cells denote the correspondence relationship between these two time series.

Traditional dynamic time warping algorithm is only applicable to univariate time series. To utilize dynamic time warping in multivariate time series and apply Mahalanobis distance, the local distance $d(i,k)$ is defined as

$$d(X^i, Y^j) = d_M(X^i, Y^j) \quad (15)$$

where $X^i$ represents the $i^{th}$ row in (2), and $Y^j$ represents the $j^{th}$ row in (3).

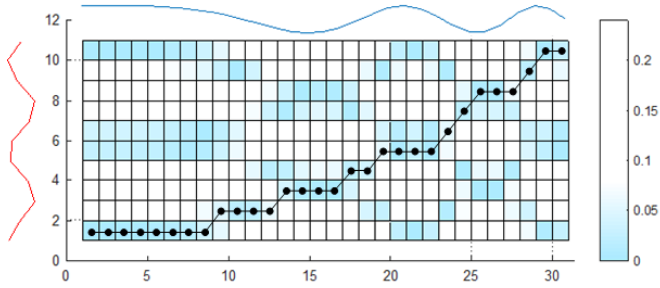

Fig.2 . An optimal warping path.

Now the distance between out-of-sync MTS is measured correctly. *K*-nearest neighbors algorithm (*k*-NN) is used for classification. The input of *k*-NN is the multivariate time series constructed by PMU measurements of generator rotor angle and active power; the output is a class membership. An object is classified by a majority vote of its neighbors, with the object being assigned to the class most common among its *k* nearest neighbors.

### III. NUMERICAL EXPERIMENTS

In this section, numerical results of the proposed method are presented. The PMU measurement data sets used in case studies are generated by time-domain simulations using *DSATools$^{TM}$*. The sampling rate is 25Hz.

A sinusoidal signal served as the forced oscillation disturbance is injected into excitation systems. Rotor angle and active power of each generator serve as features. For training data sets, the time series begins at 0s and ends at 15s, while for testing data sets, there is a delay *d* corresponding to the time needed for oscillation detection. Therefore, the testing time series starts *d* seconds later than the beginning of the oscillation and the window size of time is 5 seconds. The constructed multivariate time series is the same as (2), where *p* equals to two times the number of generators in the power grid, *h* equals to 25×15=375 for training sequences, and 25×5=125 for the testing sequence.

#### A. Four-machine Two-area Model

The detailed model parameters can be found in [15]. All generators are in a second-order model, while the load is in a constant impedance model. With the SSAT software, a detailed modal analysis on the system is conducted. The analysis results show that all natural modes have reasonably good damping and the system has a natural mode at $f_0$=0.6208Hz.

A sinusoidal signal $\Delta ref=k \times sin(2\pi ft)$ emulating the oscillation disturbance is added to the reference signal of excitation systems, where *k*=0.03.The oscillation begins at *t*=0s. For each instance, only one of the generators is injected with the oscillation disturbance. To generate enough number of valid samples, the system load varies randomly between 90 percent and 110 percent of the original load. PSAT is applied to solve the power flow equations and the infeasible load conditions are removed. In reality, system parameters are time-varying and random, which leads to the discrepancy between the model and the real system. What's more, the frequency of oscillation source can cause resonance fluctuates within a specific range. To emulate practical situation, the damping factor of each generator which influences the low-frequency oscillation most is chosen randomly in [0,4]. The frequency of oscillation source *f* fluctuates in the range of 90 percent to 110 percent of $f_0$.

Since only one of the generators acting as the oscillation source, there are four scenarios in total. Now the oscillation source locating problem converts to a multiclass classification problem. Following rules mentioned above, 800 samples (200 samples for each scenario) are generated. The ratio of training data sets and testing data sets is 1:1. To make the simulation more practical, Gaussian White Noise with the signal-to-noise ratio as 13dB is superimposed to the simulated PMU measurements. Assume the low-frequency oscillation can be detected in 3 seconds, so the testing time series begins at *d*=3s. The dimension of training and testing multivariate time series is 375-by-8, 125-by-8 respectively. In the metric learning process, dynamic triplet constraint building strategy is applied to select triplet constraints. With the obtained Mahalanobis matrix, the distance between training samples and testing samples can be calculated. After that, *k*-nearest neighbor classification is applied to select the label of the nearest trained classifier as the category of the test sample with *k*=1.

Table I presents the accuracy of oscillation source location. According to Table I, the accuracy of all four situations is 100%, which indicates the effectiveness of the proposed approach.

TABLE I PERFORMANCE TEST OF OSCILLATION SOURCE IDENTIFICATION FOR FOUR-MACHINE TWO-AREA MODEL

| Generator with disturbance | Correct | Error | Accuracy/% |
|---|---|---|---|
| $G_1$ | 100 | 0 | 100 |
| $G_2$ | 100 | 0 | 100 |
| $G_3$ | 100 | 0 | 100 |
| $G_4$ | 100 | 0 | 100 |

#### B. IEEE 39-bus System

The detailed model parameters can be found in [16]. Generators are in a fourth-order model, while the constant impedance load model is adopted. Based on the detailed modal analysis of the system, a natural mode with the frequency at $f_0$=1.3217Hz exists.

Similar to the four-machine two-area case, a sinusoidal signal $\Delta ref=k \times sin(2\pi ft)$ with *k*=0.6 is added to the reference signal of excitation systems. Since each time only one generator acting as the forced oscillation source, there are ten scenarios in total. The oscillation disturbance adds to the system at *t*=0s. Considering randomness in system load, the frequency of forced oscillation source and damping factor of each generator, 4000 samples are generated. The ratio of training data sets and testing data sets is 1:1. Gaussian White Noise with the signal-to-noise ratio equaling to 13dB is superimposed to the simulated PMU measurements. The testing time series begins at *d*=4s.

Table II presents the oscillation source location performance of proposed approach. Another machine learning approach, CELL&Decision tree approach from [8], is employed as comparison. From table II, the overall accuracy of proposal is 97.8%, while the accuracy of several scenarios reaches 100%, satisfying the accurate positioning request of engineering practice. In each scenario, the proposal outperforms the CELL&Decision method.

TABLE II PERFORMANCE TEST OF OSCILLATION SOURCE IDENTIFICATION FOR IEEE 39-BUS SYSTEM

| Generator with disturbance | Accuracy of proposal/% | Accuracy of CELL&Decision/% |
|---|---|---|
| $G_1$ | 100 | 95.7 |
| $G_2$ | 100 | 92.8 |
| $G_3$ | 100 | 98.7 |
| $G_4$ | 100 | 99.3 |
| $G_5$ | 92 | 91.3 |
| $G_6$ | 100 | 100 |
| $G_7$ | 100 | 97.3 |
| $G_8$ | 92 | 90.1 |
| $G_9$ | 94 | 93.3 |
| $G_{10}$ | 100 | 95.7 |
| Average | 97.8 | 95.4 |

*C. Influence of Oscillation Detection Delay*

Here the influence of time delay in oscillation detection is analyzed. When forced oscillation occurs, software or system operators need some time to detect it. In our work, testing time series begins at $t=d$ second to emulate the time required for oscillation detection. For IEEE 39-bus system, we have conducted simulations with different values of *d*. Fig. 3 illustrates the relationship between the location accuracy and the delay *d*. As shown in Fig. 3, as delay *d* increases, the location accuracy declines. This is because the proposed approach localizes oscillation sources mainly based on the dynamic characteristics of features in the transient process. When the oscillation tends to be stable, the characteristic of features becomes indistinct. So the sooner the oscillation is detected, the higher is the accuracy.

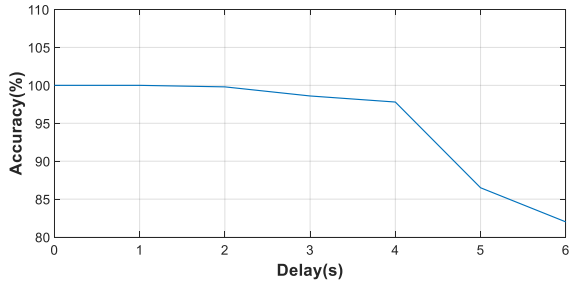

Fig.3 The relationship between location accuracy and the time delay

## IV. CONCLUSION

In this work, a multivariate time series classification method to locate the forced oscillation sources in power systems has been proposed. Mahalanobis distance of MTS constructed by rotor angle and active power of each generator is obtained through metric learning. Dynamic time warping is applied to find the optimized warping path for time series with different phases or lengths. After that, the real-time measurements can be utilized to determine the forced oscillation source by comparing the Mahalanobis distance between the testing data and the training data. Because MTS integrates all information during the transient process, it can represent the forced oscillation characteristic caused by different oscillation disturbance very well. Experimental results show the proposed approach has high location accuracy. Considering practical applications, the relationship between oscillation detection delay and detection accuracy is investigated.

The proposed approach works well for forced oscillation. In practice, poorly damped oscillation is another principal mechanism of oscillations. How to utilize the proposed approach to deal with poorly damped oscillation source location problem remains to be further investigated.